\documentstyle[12pt]{article}
\vspace{1.8cm} \oddsidemargin 0pt \evensidemargin 0pt  \topmargin
-1.5cm \footheight 26pt \footskip 10mm

\begin{document}
\baselineskip 20pt
\begin{center}
\baselineskip=24pt {\bf Extending Special Relativity to
Superluminal Motion}

\vspace{1cm} \centerline{Xiang-Yao Wu$^{a}$
\footnote{E-mail:wuxy2066@163.com}, Xiao-Jing Liu$^{a}$, Bai-Jun
Zhang$^{a}$ and Yi-Heng Wu$^{a}$}

\vskip 10pt \noindent{\footnotesize a. \textit{Institute of
Physics, Jilin Normal University, Siping 136000, China}\\
}

\end{center}
\date{}
\renewcommand{\thesection}{Sec. \Roman{section}} \topmargin 10pt
\renewcommand{\thesubsection}{ \arabic{subsection}} \topmargin 10pt
{\vskip 5mm
\begin {minipage}{140mm}
\centerline {\bf Abstract} \vskip 8pt
\par

\indent\\

\hspace{0.3in} Experiments done with single photon in the early
1990's produced a surprising result: that single photon pass
through a photon tunnel barrier with a group velocity faster than
the vacuum speed of light. Recently, a series of experiments
revealed that electromagnetic wave was able to travel at a group
velocity faster than $c$. These phenomena have been observed in
dispersive media. We think all particles can be divided into three
kinds in nature: The first kind of particle is its velocity in the
range of $0\leq v < c$, e.g. electron, atom and so on. The second
kind of particle is its velocity in the range of $0\leq v <
c_{m}$, e.g. photon. The third kind of particle is its velocity in
the range of $c\leq v < c_{m}$ ($c_{m}$ is the maximum velocity in
nature), e.g. tachyon. The first kind of particle is described by
the special relativity. In this paper, we give some new kinematic
and dynamic equations to describe the second and third kinds
particles.


\end {minipage}

\newpage
\section * {1. Introduction }

\hspace{0.3in}About one hundred years ago, Einstein laid the
foundation for a revolution in the conception of space and time,
matter and energy. Later, special theory of relativity was
accepted by mainstream physicists. It is  based on two postulates
by Einstein \cite{s1}:

1. The Principle of Relativity: All laws of nature are the same in
all inertial reference frames. In other words, we can say that the
equation expressing the laws of nature are invariant with respect
to transformations of coordinates and time from one inertial
reference frame to another.

2. The Universal Speed of Light: The speed of light in vacuum is
the same for all inertia observers, regardless of the motion of
the source, the observer, or any assumed medium of propagation.

The invariant principle of the speed of light is right in all
inertial reference frames in which their relative velocity $u$ is
less than the speed of light $c$. Since the light velocity has no
relation with the movement of light source, which has been proved
by experiment \cite{s2}, we can treat a moving light source as an
inertial reference frame, and we can obtain the result that the
speed of light has nothing to do with the moving speed of the
inertial reference frame, i.e., the speed of light is the same in
all inertial reference frames. It derives that light has no
interaction with light source, and the light has no inertia. So,
the rest mass of light tend to zero. Recently, a series of
experiments revealed that electromagnetic wave was able to travel
at a group velocity faster than $c$. These phenomena have been
observed in dispersive media [3, 4], in electronic circuits
\cite{s5}, and in evanescent wave cases \cite{s6}. In fact, over
the last decade, the discussion of the tunnelling time problem has
experienced a new stimulus by the results of analogous experiments
with evanescent electromagnetic wave packets \cite{s7}, and the
superluminal effects of evanescent waves have been revealed in
photon tunnelling experiments in both the optical domain and the
microwave range \cite{s6}. In nature, there is superluminal
phenomena, and the relative velocity $u$ between two inertial
reference frames can be larger than $c$ or equal to $c$.
Otherwise, the superluminal phenomena can also appear in the
progress of light propagation. For example, when a beam of light
moves at the same direction, the relative velocity $u$ between all
photons is equal to zero, it is not the speed of light $c$. So,
the postulation about the invariant principle of the speed of
light shouldn't be correct when the relative velocity of two
inertial reference frames is equal to the speed of light (we
regard the light as an inertial reference frames). When two beams
of light move at the opposite direction, their relative velocity
$u$ exceeds the speed of light $c$, the result that the speed of
light $c$ is the maximum speed shouldn't be correct. So, when two
inertial reference frames relative velocity $u$ is larger than the
speed of light or equal to the speed of light, Einstein's
invariant principle of the speed of light should be modified. In
nature, particles can be divided into three kinds. The first kind
of particle's velocity is in the range of $0\leq v < c$, e.g.
electron, atom and so on. The second kind of particle's velocity
is in the range of $0\leq v < c_{m}$, e.g. photon. The third kind
of particle's velocity is in the range of $c\leq v < c_{m}$
($c_{m}$ is the maximum velocity in nature), e.g. tachyon. The
first kind of particle is described by the special relativity. The
second and third kinds of particle will be researched in this
paper.

\section * {2. The space-time transformation and mass-energy relation
for the first kind of particle ($0\leq v <c$)}

The first kind of particle's velocity is in the range of $0\leq u
< c$, e.g. electron, atom and so on, and they can be described by
the special relativity. In 1905, Einstein gave the space-time
transformation and mass-energy relation which are based on his two
postulates. The space-time transformation is
\begin{eqnarray}
x&=&\frac{x^{'}+u
t^{'}}{\sqrt{1-\frac{u^{2}}{c^{2}}}}\nonumber\\
 y&=&y^{'}\nonumber\\
z&=&z^{'}\nonumber\\
t&=&\frac{t^{'}+\frac{u}{c^{2}}x^{'}}{\sqrt{1-\frac{u^{2}}{c^{2}}}},
\end{eqnarray}
where $x$, $y$, $z$, $t$ are space-time coordinates in $\sum$
frame, $x^{'}$, $y^{'}$, $z^{'}$, $t^{'}$ are space-time
coordinates in $\sum^{'}$ frame, $c$ is the speed of light, and
$u$ is the relative velocity between $\sum$ and $\sum^{'}$ frame,
which move along $x$ and $x^{'}$ axes.
The velocity transformation is\\
\begin{eqnarray}
u_{x}&=&\frac{u_{x}^{'}+u
}{1+\frac{u u_{x}^{'}}{c^{2}}}\nonumber\\
u_{y}&=&\frac{u_{y}^{'}\sqrt{1-\frac{u^{2}}{c^{2}}}}
{1+\frac{u u_{x}^{'}}{c^{2}}}\nonumber\\
u_{z}&=&\frac{u_{z}^{'}\sqrt{1-\frac{u^{2}}{c^{2}}}} {1+\frac{u
u_{x}^{'}}{c^{2}}},
\end{eqnarray}
where $u_{x}$, $u_{y}$ and $u_{z}$ are a particle velocity
projection in $\sum$ frame, $u_{x}^{'}$, $u_{y}^{'}$ and
$u_{z}^{'}$ are the particle velocity projection in $\sum^{'}$
frame. The relation between a particle mass $m$ and its velocity
$\upsilon$ is
\begin{equation}
m=\frac{m_{0}}{\sqrt{1-\frac{v^{2}}{c^{2}}}},
\end{equation}
with $m_{0}$ and $m$ being the particle rest mass and relativistic
mass respectively. The relation between a particle relativistic
energy $E$ and its relativistic mass $m$ is
\begin{equation}
E=mc^{2},
\end{equation}
and the relation of particle energy $E$ with its momentum $p$ is
\begin{equation}
E^{2}=m_{0}^{2}c^{4}+p^{2}c^{2}.
\end{equation}

\section * {3. The space-time transformation and mass-energy relation
for the second kind of particle ($0\leq \upsilon < c_{m}$)}

The second kind of particle's velocity is in the range of $0\leq
\upsilon < c_{m}$ $(c<c_{m})$, e.g. photon. Recently, a series of
experiments revealed that electromagnetic wave was able to travel
at a group velocity faster than $c$. These phenomena have been
observed in dispersive media [3, 4], in electronic circuits
\cite{s5}, and in evanescent wave cases \cite{s6}. It is about 40
years before, that O.M.P. Bilaniuk, V.K. Deshpande and E.S.G.
Sudarshan have studied the space-time relation for superluminal
reference frames within the framework of special relativity [8,
9]. They assumed that the space-time and velocity transformation
of special relativity are suitable for superluminal reference
frames. They obtained the new results that the proper length
$L_{0}$ and proper time $T_{0}$ must be imaginary so that the
measured quantities, such as length $L$ and time $T$, are real. In
this paper, we extend the framework of special relativity. We
think there is superluminal particle, but the particle velocity
can not be infinity. So, we can assume that there is a limit
velocity in nature, which is called the maximum velocity $c_{m}$.
All particle movement velocities can not exceed the maximum
velocity $c_{m}$ in arbitrary inertial reference frame. In the
velocity range of $0\leq \upsilon <c_{m}$, we give two postulates
as follows:

1. The Principle of Relativity: All laws of nature are the same in
all inertial reference frames.

2. The Universal of Maximum Velocity: There is a maximum velocity
$c_{m}$ in nature, and the $c_{m}$ is invariant in all inertial
reference frames.

From the two postulates, we can obtain the space-time
transformation and velocity transformation for the second kind of
particle ($0\leq \upsilon <c_{m}$). When we replace $c$ with
$c_{m}$, we can obtain the new transformation from the Lorentz
transformation, they are
\begin{eqnarray}
x&=&\frac{x^{'}+u
t^{'}}{\sqrt{1-\frac{u^{2}}{c_{m}^{2}}}}\nonumber\\
 y&=&y^{'}\nonumber\\
z&=&z^{'}\nonumber\\
t&=&\frac{t^{'}+\frac{u}{c_{m}^{2}}x^{'}}{\sqrt{1-\frac{u^{2}}{c_{m}^{2}}}},
\end{eqnarray}
where $x$, $y$, $z$, $t$ are space-time coordinates in $\sum$
frame, $x^{'}$, $y^{'}$, $z^{'}$, $t^{'}$ are space-time
coordinates in $\sum^{'}$ frame, $c$ is the speed of light, and
$u$ is the relative velocity between $\sum$ and $\sum^{'}$ frame,
which move along $x$ and $x^{'}$ axes.
The velocity transformation is\\
\begin{eqnarray}
u_{x}&=&\frac{u_{x}^{'}+u
}{1+\frac{u u_{x}^{'}}{c_{m}^{2}}}\nonumber\\
u_{y}&=&\frac{u_{y}^{'}\sqrt{1-\frac{u^{2}}{c_{m}^{2}}}}
{1+\frac{u u_{x}^{'}}{c_{m}^{2}}}\nonumber\\
u_{z}&=&\frac{u_{z}^{'}\sqrt{1-\frac{u^{2}}{c_{m}^{2}}}}
{1+\frac{u u_{x}^{'}}{c_{m}^{2}}},
\end{eqnarray}
where $u_{x}$, $u_{y}$, $u_{z}$ and
$v=\sqrt{u_{x}^{2}+u_{y}^{2}+u_{z}^{2}}$ $(0\leq v<c_{m})$ are a
particle velocity component and velocity amplitude in $\sum$
frame, $u_{x}^{'}$, $u_{y}^{'}$, $u_{z}^{'}$ and
$v'=\sqrt{{u'}_{x}^{2}+{u'}_{y}^{2}+{u'}_{z}^{2}}$ $(0\leq
v'<c_{m})$ are the particle velocity component projection and
velocity amplitude in $\sum^{'}$ frame. Now, We can discuss the
problem of the speed of light. For two inertial reference frames
$\sum$ and $\sum^{'}$, the $\sum^{'}$ frame is a rest frame for
light, i.e., the two reference frames relative velocity $v$ is
equal to $c$. At the time $t=0$, a beam of light is emitted from
the origin $O$. From Eq. (7), we have
\begin{equation}
u_{x}=c,
\end{equation}
then
\begin{equation}
u_{x}^{'}=0,
\end{equation}
and
\begin{equation}
u_{x}=-c,
\end{equation}
and then
\begin{equation}
u_{x}^{'}=\frac{-c-c}{1+\frac{c^{2}}{c_{m}^{2}}}=-2\frac{c
c_{m}^{2}}{c^{2}+c_{m}^{2}}>-2c.
\end{equation}
It shows that the invariant principle of light velocity is
violated in the inertial reference of light velocity movement. The
relation between a particle mass $m$ and its movement velocity
$\upsilon$ is
\begin{equation}
m=\frac{m_{0}}{\sqrt{1-\frac{\upsilon^{2}}{c_{m}^{2}}}},
\end{equation}
with $m_{0}$ and $m$ being the particle rest mass and relativistic
mass respectively. The relation between a particle relativistic
energy $E$ and its relativistic mass $m$ is
\begin{equation}
E=mc_{m}^{2},
\end{equation}
and the relation between particle energy $E$ and its momentum $p$
is
\begin{equation}
E^{2}=m_{0}^{2}c_{m}^{4}+p^{2}c_{m}^{2}.
\end{equation}
In the following, we study the nature of photon. From Eqs. (12)
and (13), we have
\begin{equation}
m_{\nu}c_{m}^{2}=\frac{m_{0}}{\sqrt{1-\frac{v^{2}}{c_{m}^{2}}}}c_{m}^{2}=h\nu,
\end{equation}
where $m_{\nu}$ is photon mass, $\nu$ is photon frequency when its
velocity is $v$ ($0\leq v<c_{m}$). When $v=0$, we have
\begin{equation}
m_{0}=\frac{h}{c_{m}^{2}}\nu_{0},
\end{equation}
where $m_{0}$ and $\nu_{0}$ are photon rest mass and rest
frequency respectively. From Eqs. (15) and (16), we have
\begin{equation}
\nu=\frac{m_{0}c_{m}^{2}}{h\sqrt{1-\frac{v^{2}}{c_{m}^{2}}}}=\frac{\nu_{0}}{\sqrt{1-\frac{v^{2}}{c_{m}^{2}}}},
\end{equation}
The Eq. (17) gives the relation of a photon's frequency with its
velocity, and the relation of a photon's movement frequency $\nu$
with rest frequency $\nu_{0}$. From the relation
\begin{equation}
\nu\lambda=v,
\end{equation}
i.e.
\begin{equation}
\frac{m_{0}c_{m}^{2}}{h}\frac{1}{\sqrt{1-\frac{v^{2}}{c_{m}^{2}}}}\lambda=v,
\end{equation}
then
\begin{equation}
\lambda=\frac{hv}{m_{0}c_{m}^{2}}\sqrt{1-\frac{v^{2}}{c_{m}^{2}}},
\end{equation}
The Eq. (20) gives the relation of a photon's wavelength with its
velocity,. From Eq. (15) and (17), we can obtain the relation
between photon mass $m$ and it velocity $v$ $(0\leq v<c_{m})$
\begin{equation}
m_{\nu}=\frac{h\nu}{c_{m}^{2}}=\frac{h\nu_{0}}{c_{m}^2\sqrt{1-\frac{v^2}{{c_m}^2}}}.
\end{equation}
From Eq. (21), we find the photon mass $m_{\nu}$ is directly
proportional to $\nu$, and we can calculate the maximum velocity
$c_{m}$ if we can measure the photon mass $m_{\nu}$ when its
frequency is $\nu$. All photon possess a finite mass and their
physical implications have been discussed by many theories and
experiments [10, 11, 12]. In Ref. [11, 12], the experiment were
made by laser, and it determined the lower limit of the photon
mass $10^{-6}eV<m_{\nu}<10^{-4}eV$, we know the laser frequency is
in the range of $8.9\times 10^{13}Hz$ $\sim$ $9.23\times
10^{14}Hz$. From Eq. (21), we can estimate the maximum velocity
$c_{m}$. It is in the rang of:
$\sqrt{\frac{6.626\times10^{-34}\times
8.9\times10^{13}}{10^{-4}\times1.6\times10^{-19}}}c\leq c_{m}=
\sqrt{\frac{h\nu}{m_{\nu}}}\leq\sqrt{\frac{6.626\times10^{-34}\times
9.23\times10^{14}}{10^{-6}\times1.6\times10^{-19}}}c$, i.e.,
$60c\leq c_{m}\leq 2000c$. In Ref. [13], the experiment measured
the superluminal velocity is $310c$. In Refs. [14-15], the
experiments measured signal velocity were $4.7c$ for the microwave
and $1.7c$ for single photon.

\section * {4. The space-time transformation and mass-velocity relation
for the third kind of particle ($c\leq \upsilon < c_{m}$)}

The third kind of particle's velocity is in the range of $c\leq
\upsilon < c_{m}$ $(c<c_{m})$, e.g. tachyon. Tachyon arises
naturally in superstring theory, and the detailed analysis from
algebraic as well as space-time diagram viewpoints is given in
[16]. There are two well-known vacuum in open bosonic string field
theory (OSFT) [17]. One is the unstable (perturbative) vacuum and
the other is the tachyon (nonperturbative) vacuum. M. Schnabl
obtained the analytic solution for the tachyon vacuum [18]. After
the Schnabl¡¯s work, there has been remarkable progress in
understanding of OSFT [19]. Especially, many works have been
devoted to the construction of analytic solutions in bosonic
string [20], and superstring field theories[21, 22].

 In the following, we will give the relation of space-time in
two inertial reference frames $\sum$ and $\sum^{'}$ for the third
kind of particle ($c\leq \upsilon < c_{m}$). We think the tachyon
travels faster than light, but its velocity can not be infinity.
So, we can assume there is a limit velocity $c_{m}$ in nature. For
the tachyon ($c\leq \upsilon <c_{m}$), we also give two postulates
as follows:

1. The Principle of Relativity: All laws of nature are the same in
all inertial reference frames.

2. The Universal of Maximum Velocity: There is a maximum velocity
$c_{m}$ in nature, and the $c_{m}$ is invariant in all inertial
reference frames.

From the two postulates, we can obtain the space-time
transformation and velocity transformation for the third kind of
particle ($c\leq \upsilon <c_{m}$). When we replace $c$ with
$c_{m}$, we can obtain the new transformation from the Lorentz
transformation, they are
\begin{eqnarray}
x&=&\frac{x^{'}+u
t^{'}}{\sqrt{1-\frac{u^{2}}{c_{m}^{2}}}}\nonumber\\
 y&=&y^{'}\nonumber\\
z&=&z^{'}\nonumber\\
t&=&\frac{t^{'}+\frac{u}{c_{m}^{2}}x^{'}}{\sqrt{1-\frac{u^{2}}{c_{m}^{2}}}},
\end{eqnarray}
where $x$, $y$, $z$, $t$ are space-time coordinates in $\sum$
frame, $x^{'}$, $y^{'}$, $z^{'}$, $t^{'}$ are space-time
coordinates in $\sum^{'}$ frame, $c$ is the speed of light, and
$u$ is the relative velocity between $\sum$ and $\sum^{'}$ frame,
which move along $x$ and $x^{'}$ axes.
The velocity transformation is\\
\begin{eqnarray}
u_{x}&=&\frac{u_{x}^{'}+u
}{1+\frac{u u_{x}^{'}}{c_{m}^{2}}}\nonumber\\
u_{y}&=&\frac{u_{y}^{'}\sqrt{1-\frac{u^{2}}{c_{m}^{2}}}}
{1+\frac{u u_{x}^{'}}{c_{m}^{2}}}\nonumber\\
u_{z}&=&\frac{u_{z}^{'}\sqrt{1-\frac{u^{2}}{c_{m}^{2}}}}
{1+\frac{u u_{x}^{'}}{c_{m}^{2}}},
\end{eqnarray}
where $u_{x}$, $u_{y}$, $u_{z}$ and
$v=\sqrt{u_{x}^{2}+u_{y}^{2}+u_{z}^{2}}$ $(c\leq v<c_{m})$ are a
particle velocity component and velocity in $\sum$ frame,
$u_{x}^{'}$, $u_{y}^{'}$, $u_{z}^{'}$ and
$v'=\sqrt{{u'}_{x}^{2}+{u'}_{y}^{2}+{u'}_{z}^{2}}$ $(c\leq
v'<c_{m})$ are the particle velocity component and velocity in
$\sum^{'}$ frame. In the following, we will give the new relation
of particle mass $m$ with its velocity $v$. We can consider the
collision between two identical particle. It is shown in Figure 1.

\setlength{\unitlength}{0.1in}
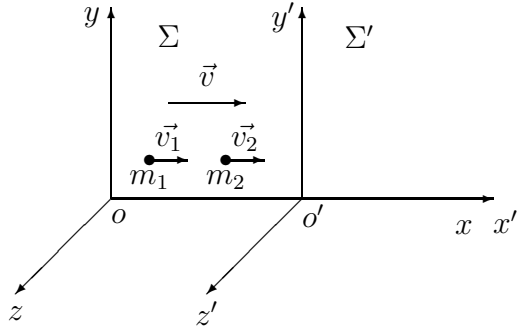
\begin{figure}
\begin{picture}(100,10)
\put(20,4){\vector(1,0){20}}
  \put(38,2){\makebox(2,1)[l]{$x$}}
  \put(40,2.3){\makebox(2,1)[l]{$x^{\prime}$}}
\put(23,9){\vector(1,0){4}}
  \put(24,10){\makebox(2,1)[c]{$\vec{v}$}}
\put(22,6){\vector(1,0){2}}
  \put(22,6.7){\makebox(2,1)[c]{$\vec{v_{1}}$}}
  \put(21,4.5){\makebox(2,1)[c]{$m_{1}$}}
\put(26,6){\vector(1,0){2}}
  \put(26,6.7){\makebox(2,1)[c]{$\vec{v_{2}}$}}
  \put(25,4.5){\makebox(2,1)[c]{$m_{2}$}}
\put(20,4){\vector(0,1){10}}
  \put(18,13){\makebox(2,1)[c]{$y$}}
\put(30,4){\vector(0,1){10}}
  \put(28,13){\makebox(2,1)[c]{$y^{\prime}$}}
\put(30,4){\vector(-1,-1){5}}
  \put(24,-2.5){\makebox(2,1)[c]{$z^{\prime}$}}
\put(20,4){\vector(-1,-1){5}}
  \put(14,-2.5){\makebox(2,1)[c]{$z$}}

  \put(30,2.6){\makebox(2,1)[l]{$o^{\prime}$}}
  \put(20,2.6){\makebox(2,1)[l]{$o$}}
  \put(32,12){\makebox(2,1)[c]{$\Sigma^{\prime}$}}
  \put(22,12){\makebox(2,1)[c]{$\Sigma$}}
 \put(22,6){\circle*{0.5}}
 \put(26,6){\circle*{0.5}}
\end{picture}

 \vskip 10pt

\caption{The $\Sigma$ is the laboratory system, $\Sigma^{\prime}$
is the mass-center system for two particles $m_1$ and $m_2$, and
the two inertial frames relative velocity is $v$.} \label{moment}
\end{figure}

The $\Sigma$ is the laboratory system, and $\Sigma^{\prime}$ is
the mass-center system of two particles $m_{1}$ and $m_{2}$. In
$\Sigma$ system, the velocity of two particles $m_{1}$ and $m_{2}$
are $\vec{v_{1}}$ and $\vec{v_{2}}$ $(c\leq v_{2}< v_{1}<c_m)$,
which are along with $x (x^{\prime})$ axis, and they are
$v^{\prime}$ and $-v^{\prime}$ in $\Sigma^{\prime}$ system. After
collision, the two particles velocities are all $v$ $(c\leq
v<c_m)$ in $\Sigma$ system. The momentum was conserved in this
process:
\begin{equation}
m_{1}v_{1}+m_{2}v_{2}=(m_{1}+m_{2})v.
\end{equation}
According to equation (23),
\begin{eqnarray}
v_{1}=\frac{v^{\prime}+v}{1+\frac{v v^{\prime}}{c_{m}^{2}}} \nonumber\\
v_{2}=\frac{-v^{\prime}+v}{1-\frac{v v^{\prime}}{c_{m}^{2}}}.
\end{eqnarray}
From equations (24) and (25), we get
\begin{equation}
m_{1}(1-\frac{v v^{\prime}}{c_{m}^{2}})=m_{2}(1+\frac{v
v^{\prime}}{c_{m}^{2}}),
\end{equation}
from equation (23), we can obtain
\begin{equation}
1-\frac{u_{x}^{\prime}u}{c_{m}^{2}}=\frac{\sqrt{1-\frac{{v}^{
2}}{c_{m}^{2}}} \sqrt{1-\frac{u^{2}}{c_{m}^{2}}}}
{\sqrt{1-\frac{{v'}^{2}}{c_{m}^{2}}}},
\end{equation}
where $v=\sqrt{u_{x}^{2}+u_{y}^{2}+u_{z}^{2}}$ and
$v'=\sqrt{{u'}_{x}^{2}+{u'}_{y}^{2}+{u'}_{z}^{2}}$. For the
particle $m_{1}$, the equation (27) becomes
\begin{equation}
1+\frac{v^{\prime}v}{c_{m}^{2}}=\frac{\sqrt{1-\frac{v^{\prime2}}{c_{m}^{2}}}
\sqrt{1-\frac{v^{2}}{c_{m}^{2}}}}
{\sqrt{1-\frac{v_{1}^{2}}{c_{m}^{2}}}}.
\end{equation}
For the particle $m_{2}$, the equation (27) becomes
\begin{equation}
1-\frac{v^{\prime}v}{c_{m}^{2}}=\frac{\sqrt{1-\frac{v^{\prime2}}{c_{m}^{2}}}
\sqrt{1-\frac{v^{2}}{c_{m}^{2}}}}
{\sqrt{1-\frac{v_{2}^{2}}{c_{m}^{2}}}}.
\end{equation}
By substituting equations (28) and (29) into (26), we get
\begin{equation}
m(v_{1})\sqrt{1-\frac{v_{1}^{2}}{c_{m}^{2}}}=
m(v_{2})\sqrt{1-\frac{v_{2}^{2}}{c_{m}^{2}}}=
m(c)\sqrt{1-\frac{c^{2}}{c_{m}^{2}}}=constant,
\end{equation}
where $m(c)$ is the particle mass when its velocity is equal to
$c$.\\
For velocity $v$ ($c\leq v<c_m$), we have
\begin{equation}
m(v)\sqrt{1-\frac{v^{2}}{c_{m}^{2}}}=m(c)\sqrt{1-\frac{c^{2}}{c_{m}^{2}}},
\end{equation}
and hence
\begin{equation}
m(v)=m_{c}\sqrt{\frac{c_{m}^{2}-c^{2}}{c_{m}^{2}-v^{2}}}.
\end{equation}
with $m_{c}=m(c)$. The equation (32) is the relation between
tachyon mass $m$ and its velocity $v$ ($c\leq v<c_m$).

\section * {5. The relation of energy with mass for the second and third kinds particles}

\hspace{0.3in}For the first kind particle ($0\leq v<c$), the
4-vector is defined
\begin{equation}
x_{\mu}=(x_{1}, x_{2}, x_{3}, x_{4})=(x, y, z, ict).
\end{equation}
The invariant interval $ds^{2}$ is given by the equation
\begin{equation}
ds^{2}=-dx_{\mu}dx_{\mu}=c^{2}dt^{2}-(dx)^{2}-(dy)^{2}-(dz)^{2}=c^{2}d\tau^{2},
\end{equation}
we get
\begin{equation}
d\tau=\frac{1}{c}ds,
\end{equation}
where $d\tau$ is the proper time, the 4-velocity can be defined by
\begin{equation}
U_{\mu}=\frac{dx_{\mu}}{d\tau}=\frac{dx_{\mu}}{dt}\frac{dt}{d\tau}=\gamma_{\mu}(\vec{v},
ic),
\end{equation}
where $\gamma_{\mu}=\frac{1}{\sqrt{1-\frac{v^{2}}{c^{2}}}}$,
$\vec{v}=\frac{d\vec{x}}{dt}$ and $\frac{dt}{d\tau}=\gamma_{\mu}$.
We can define the 4-momentum as
\begin{equation}
p_{\mu}=m_{0}U_{\mu}=(\vec{p}, ip_{4}),
\end{equation}
with $\vec{p}=\frac{m_{0}\vec{v}}{\sqrt{1-\frac{v^{2}}{c^{2}}}}$,
$p_{4}=\frac{m_{0}c}{\sqrt{1-\frac{v^{2}}{c^{2}}}}$. We can define
the energy of a particle as
\begin{equation}
E=\frac{m_{0}c^{2}}{\sqrt{1-\frac{v^{2}}{c^{2}}}},
\end{equation}
then
\begin{equation}
p_{\mu}=(\vec{p}, \frac{i}{c}E),
\end{equation}
the invariant quantity constructed from this 4-vector is
\begin{equation}
p_{\mu}p_{\mu}=p^{2}-\frac{E^{2}}{c^{2}}=-m_{0}^{2}c^{2},
\end{equation}
i.e.,
\begin{equation}
E^{2}-p^{2}c^{2}=m_{0}^{2}c^{4}.
\end{equation}
Eq. (41) is the relation among a particle mass $m$, momentum
$\vec{p}$ and energy $E$.\\

For the second kind particle ($0\leq v< c_{m}$), when $c$ is
replaced with $c_m$, we can easily obtain the relation between
energy and mass, and easily obtain the relation among a particle
mass $m$, momentum $\vec{p}$ and energy $E$ from Eqs. (33)-(41).
\begin{equation}
E=\frac{m_{0}c_{m}^{2}}{\sqrt{1-\frac{v^{2}}{c_{m}^{2}}}},
\end{equation}
and
\begin{equation}
E^{2}-p^{2}c^{2}=m_{0}^{2}c_{m}^{4}.
\end{equation}
\hspace{0.3in}For the third kind particle ($c\leq v< c_{m}$), when
$c(m_0)$ is replaced with $c_m(m_c)$ we can also obtain the
relation between energy and mass, and easily obtain the relation
among a particle mass $m$, momentum $\vec{p}$ and energy $E$ from
Eqs. (33)-(41).
\begin{equation}
E=\frac{m_{c}c_{m}^{2}}{\sqrt{1-\frac{v^{2}}{c_{m}^{2}}}},
\end{equation}
and
\begin{equation}
E^{2}-p^{2}c_{m}^{2}=m_{c}^{2}c_{m}^{4}.
\end{equation}
By substituting Eq. (32) into (44), we can obtain the relation of
mass-energy for the tachyon ($c\leq v< c_{m}$).
\begin{equation}
E=\frac{m(v)}{\sqrt{c_{m}^{2}-c^{2}}}c_{m}^{3}.
\end{equation}

\section * {6. The relativistic dynamics for the second and third kinds particles}

For the first kind particle ($0\leq v<c$), we know the 4-force is
defined as:
\begin{equation}
K_{\mu}=\frac{dp_{\mu}}{d\tau}.
\end{equation}
From equation (39), we have
\begin{equation}
K_{\mu}=(\vec{K}, iK_{4}),
\end{equation}
the "ordinary" force $\vec{K}$ is
\begin{equation}
\vec{K}=\frac{d\vec{p}}{dt}
\frac{dt}{d\tau}=\frac{1}{\sqrt{1-\frac{v^{2}}{c^{2}}}}\frac{d\vec{p}}{dt},
\end{equation}
while the fourth component
\begin{eqnarray}
K_{4}&=&\frac{dp_{4}}{d\tau}=\frac{1}{c}\frac{dE}{d\tau} \nonumber\\
&=&\frac{1}{c}\vec{v}\cdot \vec{K},
\end{eqnarray}
and so
\begin{equation}
K_{\mu}=(\vec{K}, \frac{i}{c}\vec{v}\cdot \vec{K}),
\end{equation}
the covariant equation for a particle are
\begin{equation}
\vec{K}=\frac{d\vec{p}}{d\tau},
\end{equation}
\begin{equation}
\vec{K}\cdot \vec{v}=\frac{dE}{d\tau}.
\end{equation}
\begin{equation}
\sqrt{1-\frac{v^{2}}{c^{2}}}\vec{K}\cdot \vec{v}=\frac{dE}{dt},
\end{equation}
we define force $\vec{F}$ as
\begin{equation}
\vec{F}=\sqrt{1-\frac{v^{2}}{c^{2}}}\vec{K}.
\end{equation}
From Eqs. (52)-(54), we have the relativistic dynamics equations
for a particle are
\begin{equation}
\vec{F}=\frac{d\vec{p}}{dt},
\end{equation}
\begin{equation}
\vec{K}\cdot \vec{v}=\frac{dE}{dt}.
\end{equation}
\hspace{0.3in}For the second kind particle ($0\leq v<c_{m}$), the
relativistic dynamics equations are Eqs. (55) and (56), but some
physical
quantities should be modified as follows: \\
The 4-momentum and 4-force are
\begin{equation}
p_{\mu}=m_{0}U_{\mu}=(\vec{p}, \frac{i}{c_{m}}E),
\end{equation}
and
\begin{equation}
K_{\mu}=(\vec{K}, iK_{4}),
\end{equation}
with
$\vec{p}=\frac{m_{0}\vec{v}}{\sqrt{1-\frac{v^{2}}{c_{m}^{2}}}}$,
$p_{4}=\frac{m_{0}c_{m}}{\sqrt{1-\frac{v^{2}}{c_{m}^{2}}}}$,
$\vec{K}=\frac{d\vec{p}}{dt}\frac{1}{\sqrt{1-\frac{v^{2}}{c_{m}^{2}}}}$
and
$K_{4}=\frac{1}{c_{m}}\vec{v}\cdot \vec{K}.$\\
\hspace{0.3in}For the third kind particle ($c\leq v<c_{m}$), the
relativistic dynamics equations are also Eqs. (55) and (56), and
some physical
quantities also should be modified as follows: \\
The 4-momentum and 4-force are
\begin{equation}
p_{\mu}=m_{c}U_{\mu}=(\vec{p}, ip_{4}),
\end{equation}
and
\begin{equation}
K_{\mu}=(\vec{K}, iK_{4}),
\end{equation}
with
$\vec{p}=\frac{m_{c}\vec{v}}{\sqrt{1-\frac{v^{2}}{c_{m}^{2}}}}$,
$p_{4}=\frac{m_{c}c_{m}}{\sqrt{1-\frac{v^{2}}{c_{m}^{2}}}}$,
$\vec{K}=\frac{d\vec{p}}{dt}\frac{1}{\sqrt{1-\frac{v^{2}}{c_{m}^{2}}}}$
and $K_{4}=\frac{1}{c_{m}}\vec{v}\cdot \vec{K}$.

\section * {7. The quantum wave equation for the second and third kinds particles}

For the first kind particle ($0\leq v<c$), we express $E$ and
$\vec{p}$ as operators:
\begin{eqnarray}
E\rightarrow i\hbar\frac{\partial}{\partial t} \nonumber\\
\vec{p}\rightarrow -i\hbar \nabla,
\end{eqnarray}
we can obtain the quantum wave equation of spin $0$ particle from
Eq. (41)
\begin{equation}
[\frac{\partial^{2}}{\partial
t^{2}}-c^{2}\nabla^{2}+\frac{m_{0}^{2}c^{4}}{\hbar^{2}}]\Psi(\vec{r},
t)=0,
\end{equation}
and we can obtain the quantum wave equation of spin $\frac{1}{2}$
particle
\begin{equation}
i\hbar \frac{\partial}{\partial t}\Psi=[-i\hbar c
\vec{\alpha}\cdot \vec{\nabla}+m_{0}c^{2}\beta]\Psi,
\end{equation}
where $\alpha$ and $\beta$ are matrixes
\[
\alpha= \left (
\begin {array} {cc}
0 & \vec{\sigma} \\
\vec{\sigma} & 0
\end{array} \right ),
\]
and
\[
\beta= \left (
\begin {array} {cc}
I & 0 \\
0 & -I
\end{array} \right ),
\]
where $\vec{\sigma}$ are Pauli matrixes, and $I$ is unit matrix of
$2\times 2$.\\
For the second kind particle ($0\leq v<c_{m}$), we can obtain the
quantum wave equation of spin $0$ particle from Eq. (43)
\begin{equation}
[\frac{\partial^{2}}{\partial
t^{2}}-c_{m}^{2}\nabla^{2}+\frac{m_{0}^{2}c_{m}^{4}}{\hbar^{2}}]\Psi(\vec{r},
t)=0,
\end{equation}
and we can obtain the quantum wave equation of spin $\frac{1}{2}$
particle
\begin{equation}
i\hbar \frac{\partial}{\partial t}\Psi=[-i\hbar c_{m}
\vec{\alpha}\cdot \vec{\nabla}+m_{0}c_{m}^{2}\beta]\Psi.
\end{equation}
For the third kind particle ($c\leq v<c_{m}$), we can obtain the
quantum wave equation of spin $0$ particle from Eq. (45)
\begin{equation}
[\frac{\partial^{2}}{\partial
t^{2}}-c_{m}^{2}\nabla^{2}+\frac{m_{c}^{2}c_{m}^{4}}{\hbar^{2}}]\Psi(\vec{r},
t)=0,
\end{equation}
and we can obtain the quantum wave equation of spin $\frac{1}{2}$
particle
\begin{equation}
i\hbar \frac{\partial}{\partial t}\Psi=[-i\hbar c_{m}
\vec{\alpha}\cdot \vec{\nabla}+m_{c}c_{m}^{2}\beta]\Psi.
\end{equation}

\section * {8. Conclusion}

In nature, we think particles can be divided into three kinds. The
first kind of particle is its velocity is in the range of $0\leq v
< c$, e.g. electron, atom, macroscopical matter and so on. The
second kind of particle is its velocity in the range of $0\leq v <
c_{m}$, e.g. photon. The third kind of particle is its velocity in
the range of $c\leq v < c_{m}$ ($c_{m}$ is the maximum velocity in
nature), e.g. tachyon. The first kind of particle is described by
the special relativity. In this paper, we give the new space-time
transformation, the relation between mass and energy, the classic
dynamics equation and the quantum wave equations for the second
and third kinds particles.

\end{document}